\documentclass[preprint,aps,tightenlines,showpacs]{revtex4}
\def\be{\begin{equation}}
\def\ee{\end{equation}}
\def\bea{\begin{eqnarray}}
\def\eea{\end{eqnarray}}

\usepackage{bm}
\usepackage{graphicx}
\usepackage{subfigure}
\usepackage{epsfig}
\usepackage{enumerate}

\def\slash#1{\not\!#1}

\begin{document}

{  % \tighten

\title{Study of $K^0 \rightarrow \pi^- e^+ \nu _e  e^+ e^- $
in chiral perturbation theory}
\author
{ K. Tsuji and  T. Sato}
\affiliation
{Department of Physics, Osaka University, Toyonaka, Osaka 560-0043, Japan}

\date{\today}

\begin{abstract}
$K^0 \rightarrow \pi^- e^+ \nu_e e^+ e^-$
decay rates are studied up to the next-to-leading order(${\mathcal O}^{(4)}$)
in  chiral perturbation theory.
It is found that the ${\mathcal O}^{(4)}$ terms appreciably modify
the shape of the invariant mass distribution of
leptons($3e\nu$) and the  energy spectrum of neutrinos.
\end{abstract}


\maketitle
}

%------------------------------------------------------------

\section{introduction}

The radiative semileptonic kaon decay,
$K_L \rightarrow \pi^\pm e^\mp \nu \gamma$
($K_{l3\gamma}$)  has been studied extensively
\cite{ktev-2,ktev-3,na48-2,fearing-1,holstein,
 bijnens-2,gasser-2,kubis,moulson,bijnens}
within the chiral perturbation theory (ChPT)\cite{gasser-1}.
The amplitude of $K_{l3\gamma}$
can be written as a sum of an inner bremstrahlung(IB) amplitude
and a structure dependent (SD) amplitude.
The IB amplitude is the leading
  ${\mathcal O}(q^{-1})$ and ${\mathcal O} (q^0)$ term
of the photon momentum($q$) expansion of the amplitude and hence can
be
 related to the nonradiative $K_{l3}$ amplitude by
using the theorems of  Low\cite{low} and
 Adler and Dothan\cite{adler}.
On the other hand, the SD amplitude, which is  of the order of
${\mathcal O}(q)$ and higher, contains new information  on the
hadron currents and therefore is the main interest for studying the
$K_L \rightarrow \pi^\pm e^\mp \nu \gamma (K_{l3\gamma})$ reaction.

Fearing {\it et al.}\cite{fearing-1}  studied the radiative $K_{l3}$
decay using the Low and Adler-Dothan theorems and the partial
conservation of the axial current (PCAC) hypothesis. Later,
Holstein\cite{holstein} analyzed this decay process using the model
independent ChPT at ${\mathcal O} (p^4)$ at tree level. The full
${\mathcal O}(p^4)$  ChPT analysis including the loop effects has
been done by Bijnens {\it et al.}~\cite{bijnens-2}. Further ChPT
analysis of the $K_{e3\gamma}$ decay up to the  ${\mathcal O}(p^6)$
terms was reported  by Gasser {\it et al.} \cite{gasser-2, kubis}.
In these calculations, the effect of the SD was found to be rather
small in determining the integrated decay rate, but has appreciable
effects
 on the differential decay rates.
Sizable effects of the SD amplitude are found on the photon energy
spectrum and pion energy distribution in the kinematic region where
the experimental counting rate is small. The comparison of the data
and a review of the theoretical studies on this reaction are given
in Ref.  \cite{moulson} and Ref. \cite{bijnens}, respectively.

In this paper, we report on
a ChPT study  of the semileptonic decay process of kaon
 $K^0 \rightarrow \pi^- e^+ \nu_e e^+ e^-$
($K^0_{e3e^+e^-}$) process which differs from the $K_L \rightarrow
\pi^\pm e^\mp \nu \gamma$ ($K_{l3\gamma}$)discussed above because it
involves the production of a timelike virtual photon followed by
its decay into a $e^+e^-$ pair. The recent KTeV
experiment\cite{kotera} is capable of measuring various differential
decay rates of this reaction. 
The measured invariant mass distributions
of $e^+e^-$($M_{e^+e^-}$) and leptons($M_{3e\nu}$)
 and the energy spectrum of the neutrino
will provide information for testing the ChPT
predictions and extracting the information on the hadronic
matrix elements. 
To analyze the data from this experiment and
similar future experiments, it is necessary to have a full
${\mathcal O}(p^4)$ ChPT prediction of the differential decay rates
of the $K^0_{e3e^+e^-}$ process. The purpose of this paper is to
carry out such a calculation which, to our knowledge, is currently
not available. We will examine various invariant mass distributions
and energy spectrum. In particular we study which observables are
more sensitive to the ${\mathcal O}(p^4)$ terms. We also predict the
branching ratios of $K^0_{l3e^+e^-}(l=e,\mu)$ decay relative to
$K^0_{l3}$ decay. We however have not extracted the effects of the
SD term and leave this more difficult problem for future
investigations.

In section II, we summarize the effective Lagrangian employed in
this work. The explicit form of the amplitudes of
$K^0_{l3e^+e^-}(l=e,\mu)$ from ChPT up to ${\mathcal O}(p^4)$ are
presented in Section III. The invariant mass spectrum of the
$K^0_{l3e^+e^-}(l=e,\mu)$ decay and effects of the ${\mathcal
O}(p^4)$ amplitudes are discussed in section IV.

\section{Effective Lagrangian}

Chiral perturbation theory is an effective field theory of QCD
to describe low energy hadronic system using a systematic
perturbation scheme.  In this section, a standard effective
Lagrangian of ChPT for Goldstone bosons~\cite{gasser-1,scherer}
 is summarized for completeness.
The dynamical variable of ChPT, $U(x)$, is parametrized by using
octet fields $\phi_a$ of Goldstone bosons  as
\begin{equation}
U = \exp\left( i \frac{\phi (x)}{F_0} \right), \qquad
\end{equation}
with
\begin{equation}
\phi (x) = \sum_{a=1}^8\lambda_a \phi_a(x).
\end{equation}
Here $F_0$ is the pion decay constant in the chiral limit and
$\lambda_a$ are the Gell-Mann matrices. Following the standard
counting rule~\cite{gasser-1,scherer}, the leading  ${\mathcal
O}(p^2)$  order chiral effective Lagrangian is given as
\begin{equation}
{\mathcal L}^{(2)}  = \frac{F_0^2}{4} <D_\mu U
(D^\mu U)^\dagger
+ \chi U^\dagger + U\chi ^\dagger>.\label{001}
\end{equation}
$<O>$ denotes the trace of the matrix $O$. $\chi$ is given as  $\chi
= 2B_0M$ using the quark mass matrix $M =
\textrm{diag}(m_u,m_d,m_s)$ and parameter $B_0$. The covariant
derivative $D_\mu U$, which includes a external electromagnetic
field ($A_\mu$) and the charged weak boson ($W_\mu$), is defined as
follows,
\begin{eqnarray}
D_\mu U &=& \partial_\mu U - ir_\mu U + iUl_\mu ,\\
r_\mu &=& v_\mu + a_\mu = -eQA_\mu ,\\
l_\mu &=& v_\mu - a_\mu = -eQA_\mu -\frac{g}{\sqrt{2}}
(W_\mu^+T_+ + \textrm{h.c.}).
\end{eqnarray}
$Q$ is the quark charge and $T_+$ is given by the
CKM matrix elements.
\begin{equation}
Q=\frac{1}{3}
\left(
\begin{array}{ccc}
2 & 0 & 0 \\
0 &-1 & 0 \\
0 & 0 & -1 
\end{array}
\right), \qquad
T_+ = 
\left(
\begin{array}{ccc}
0 & V_{ud} & V_{us} \\
0 & 0 & 0 \\
0 & 0 & 0
\end{array}
\right).
\end{equation}

The next to-leading-order (NLO) ${\mathcal O}(p^4)$ effective Lagrangian
following Ref. \cite{gasser-1} is given as
\begin{eqnarray}
\mathcal{L}^{(4)} &=&  L_1< D_\mu U   (D^\mu U) ^\dagger >^2 
+L_2<D_\mu U  (D_\nu U)^\dagger  ><D^\mu U  (D^\nu U)^\dagger > \nonumber \\
&&+L_3<D_\mu U (D^\mu U)^\dagger  D_\nu U  (D^\nu U)^\dagger  >  
+L_4< D_\mu U   (D^\mu U)^\dagger ><\chi U^\dagger + U\chi^\dagger>\nonumber \\
&&+L_5< D_\mu U   (D^\mu U)^\dagger (\chi U^\dagger + U\chi^\dagger)>
+L_6<\chi U^\dagger + U\chi ^\dagger>^2 \nonumber \\
&&+L_7<\chi^\dagger U - U^\dagger \chi>^2 
+L_8 <\chi U^\dagger \chi U^\dagger + U\chi ^\dagger U \chi ^\dagger >\nonumber\\
&&-iL_9<f_{\mu \nu}^RD^\mu U (D^\nu U)^\dagger 
+ f_{\mu \nu}^L D^\mu D  (D^\nu U)^\dagger > 
+L_{10}<Uf^L_{\mu \nu}U^\dagger f_R^{\mu \nu}>, \label{002}
\end{eqnarray}
with
\begin{eqnarray}
f^{R(L)}_{\mu \nu} & = &\partial_\mu r(l)_\nu 
- \partial_\nu r(l)_\mu - i[r(l)_\mu, r(l)_\nu],
\end{eqnarray}
and its filed tensors are defined as
\begin{eqnarray}
f^R_{\mu\nu} &=& \partial_\mu r_\nu - \partial_\nu r_\mu - i[r_\mu, r_\nu],\\
f^L_{\mu\nu} &=& \partial_\mu l_\nu - \partial_\nu l_\mu - i[l_\mu, l_\nu].
\end{eqnarray}

The following piece of the chiral anomaly term~\cite{wz,witten}
contributes to the ${\mathcal O}(p^4)$ amplitude of $K^0_{e3e^+e^-}$
decay,
\begin{eqnarray}
{\mathcal L}^{(4)}_{anom}
&=& - \frac{ieg}{16\sqrt{2}\pi^2}
\epsilon^{\mu \nu \rho \sigma}
W^+_\mu
\partial_\nu A_\rho
\bigg< T_+\left\{ 
\partial_\sigma U^\dagger QU 
- 2 U^\dagger \partial_\sigma U Q
- 2 Q U^\dagger \partial_\sigma U
- U^\dagger Q \partial_\sigma U \right\}\bigg>
\nonumber \\
&& + \textrm{h.c.}.
\end{eqnarray}

\section{LO and NLO amplitudes of $K^0_{e3e^+e^-}$ }

Using the chiral effective Lagrangian presented in the previous section,
we study the amplitude of $K^0_{e3e^+e^-}$ decay 
 up to the next-to-leading order.
\begin{equation}
K^0(p_1) \rightarrow \pi^-(p_2) + \,e^+ (k_1)\, + 
\nu_e (k_2) \, + e^+ (k_3) \, + e^- (k_4). \label{reaction}
\end{equation}
The momentum of the virtual photon $q^\mu$ is given by 
$q^\mu = k_3^\mu + k_4^\mu$.
The leading order amplitude $T^{(2)}$ is obtained from
tree diagrams with vertices from ${\mathcal L}^{(2)}$. 
Loop diagrams from ${\mathcal L}^{(2)}$ and tree diagrams
with vertices from ${\mathcal L}^{(4)}$ generate the next-to-leading
order amplitude $T^{(4)}$.

\begin{figure}[h]
\begin{center}
\subfigure[]{
\includegraphics*[scale = .7]{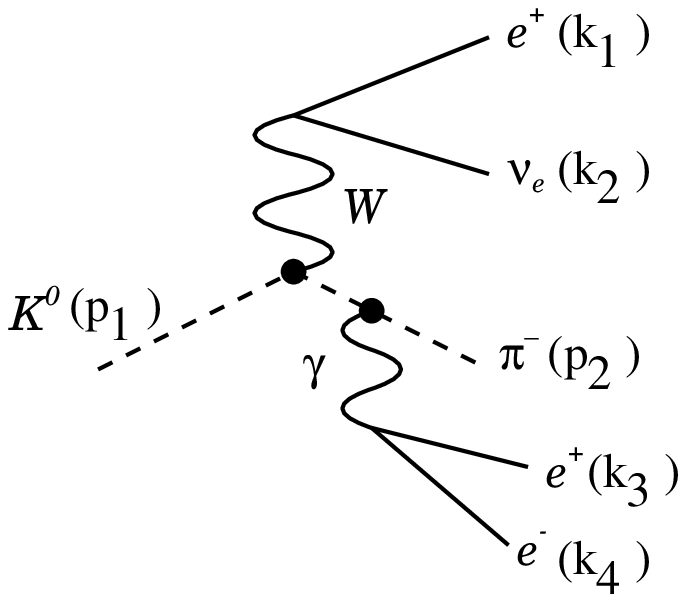}
\label{fig0-a}
}
\hspace{.5cm}
\subfigure[]{
\includegraphics*[scale = .7]{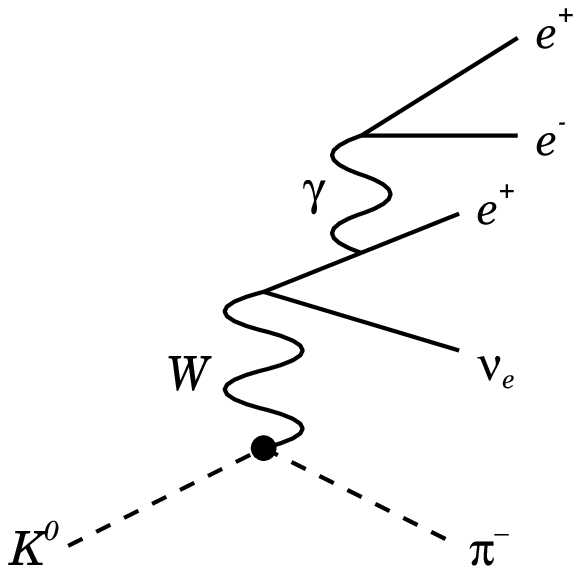}
\label{fig0-b}
}
\hspace{.5cm}
\subfigure[]{
\includegraphics*[scale = .7]{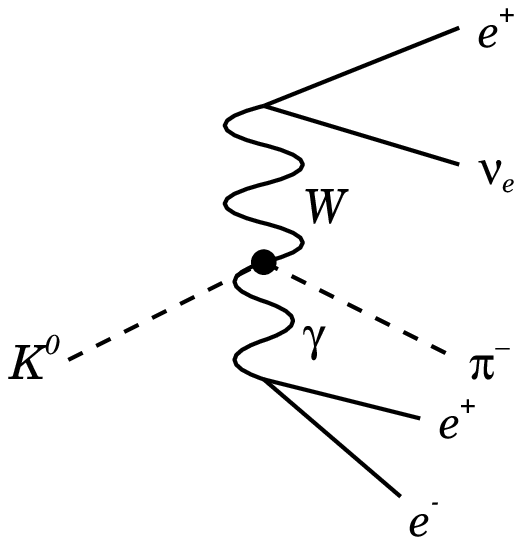}
\label{fig0-c}
}
\caption{The leading order diagrams
contributing to the $K^0_{e3e^+e^-}$ decay.
The dark circles are LO  vertices from ${\mathcal L}^{(2)}$.}
\label{figure0}
\end{center}
\end{figure}

The leading order amplitude $T^{(2)}$ of the $K^0_{e3e^+e^-}$ decay
shown in Fig. \ref{figure0} is given as
\begin{eqnarray}
T^{(2)} &=& - \frac{G_F}{\sqrt{2}}e^2V_{us}^*
\frac{1}{q^2}\bigg[
\bar{u}(k_2)\left\{ g_{\mu \nu } -
\frac{2q_\nu p_{2\mu}}
{(p_2 + q)^2 -m_\pi ^2}\right\} 
\gamma ^\nu (1-\gamma _5) v(k_1) \nonumber \\
&&+ \bar{u}(k_2)(\slash{p_1} + \slash{p_2})(1-\gamma_5)
\left\{ \frac{2k_{1\mu} + \slash{q} \gamma_\mu 
}{(k_1 + q)^2 -m_e^2}
-\frac{2p_{2\mu }}{(p_2+q)^2 - m_\pi ^2 } \right\}v(k_1)
\bigg]\bar{u}(k_4)\gamma ^\mu v(k_3). \nonumber \\
\label{leading}
\end{eqnarray}
Here $G_F$ is Fermi constant and $V_{us}$ is the CKM matrix element.
Since we have two positrons in the final state, Eq. (\ref{leading})
represents the 'direct amplitude'. The 'exchange amplitude' is given
from Eq. (\ref{leading}) by interchanging momentum and spins of the
two positrons and by taking into account the phase $(-1)$. Eq.
(\ref{leading}) satisfies gauge invariance and agrees with Eq.
(5.12) of Ref. \cite{bijnens-2} when we replace the $e^+e^-$ current
by the photon polarization vector as
\begin{equation}
\frac{e}{q^2}\bar{u}(k_4)\gamma ^\mu v(k_3) 
= \epsilon^{* \mu}.
\end{equation}

\begin{figure}[h]
\begin{center}
\subfigure[]{
\includegraphics*[scale = .7]{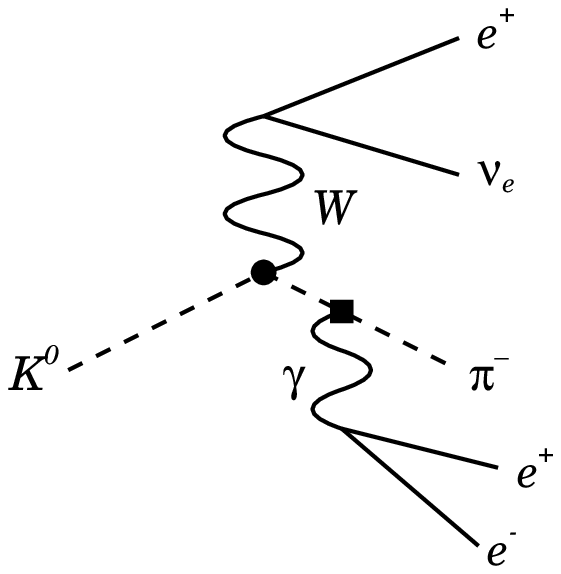}
\label{fig1-1}
}
\hspace{.5cm}
\subfigure[]{
\includegraphics*[scale = .7]{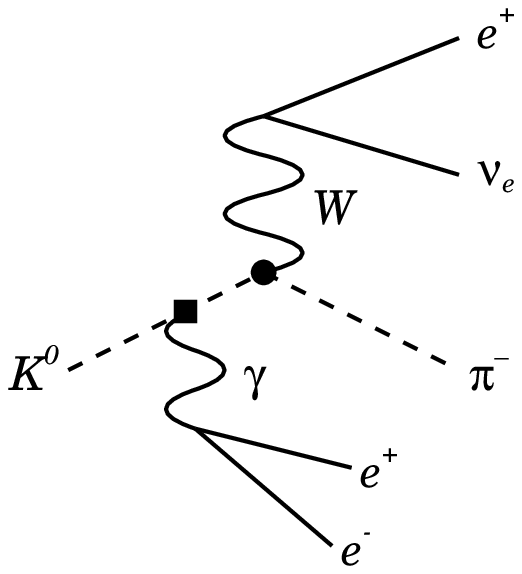}
\label{fig1-2}
}
\hspace{.5cm}
\subfigure[]{
\includegraphics*[scale = .7]{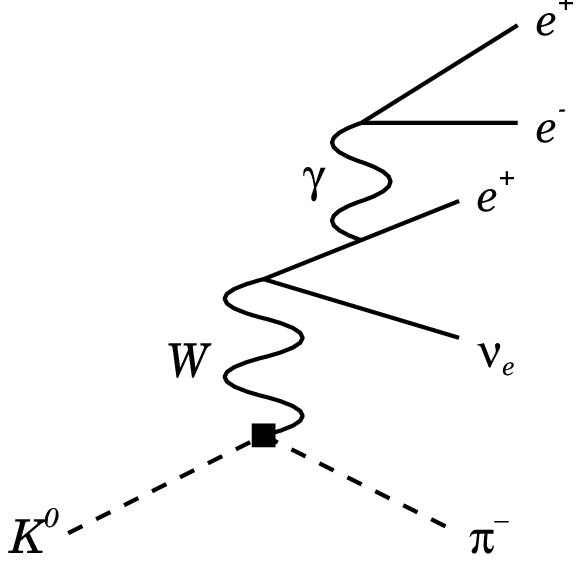}
\label{fig1-3}
}
\hspace{.5cm}
\subfigure[]{
\includegraphics*[scale = .7]{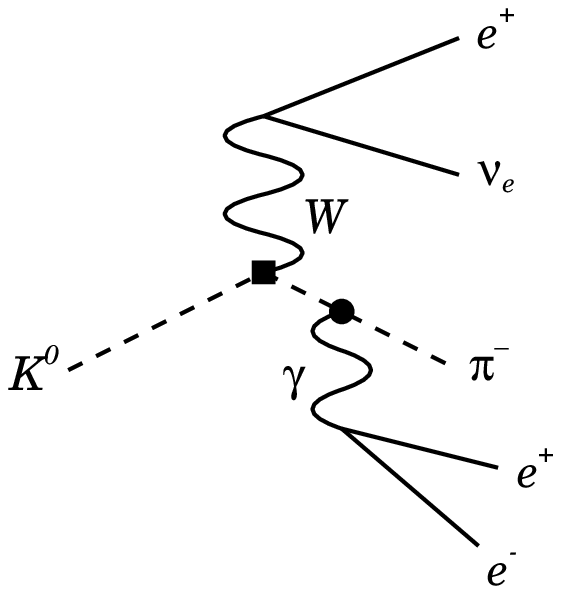}
\label{fig1-4}
}
\hspace{.5cm}
\subfigure[]{
\includegraphics*[scale = .7]{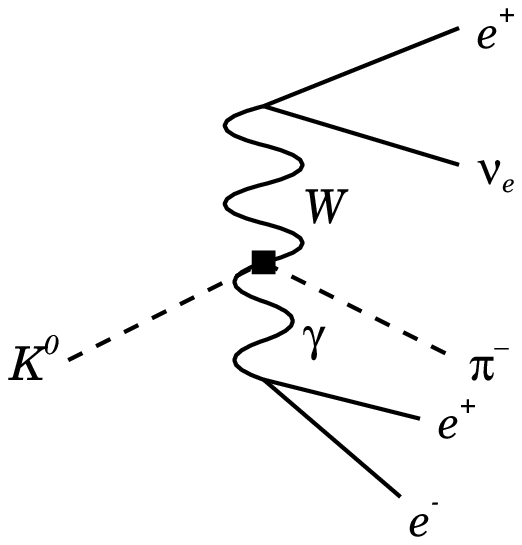}
\label{fig1-5}
}
\caption{The NLO diagrams contributing to the $K^0_{e3e^+e^-}$ decay.
The dark squares(circles) are NLO (LO) vertices. }
\label{figure1}
\end{center}
\end{figure}

At the NLO, loop corrections and contributions of $\mathcal{L}^{(4)}$
are included. We take into account the diagrams shown in Fig.
\ref{figure1}. They are the NLO correction of the pion(Fig.
\ref{fig1-1}) and  kaon(Fig. \ref{fig1-2}) electromagnetic form
factors, $\pi K W$ vertex(Fig. \ref{fig1-3} and Fig. \ref{fig1-4}),
$K\pi W \gamma$ vertex(Fig. \ref{fig1-5}) and anomaly term(Fig.
\ref{fig1-5}). The NLO T-matrix  ($T^{(4)}$) is given by the sum of
six amplitudes as,
\begin{equation}
T^{(4)} = T^{(4)}_{(a)} + T^{(4)}_{(b)}
+ T^{(4)}_{(c)} + T^{(4)}_{(d)} + T^{(4)}_{(e)}
+ T^{(4)}_{(e, anom)}.
\end{equation}
For completeness, the explicit forms of  $T^{(4)}_{(i)}$ are 
described in the next subsections. It is worth noting that, using
the formulas given in the Appendix B, the expressions of  $T^{(4)}$
can be shown to agree with those  of Ref. \cite{bijnens-2} 
 for the real photon limit.

\subsection{Pion and neutral kaon form factor}

\begin{figure}[h]
\begin{center}
\subfigure[]{
\includegraphics*[scale = .8]{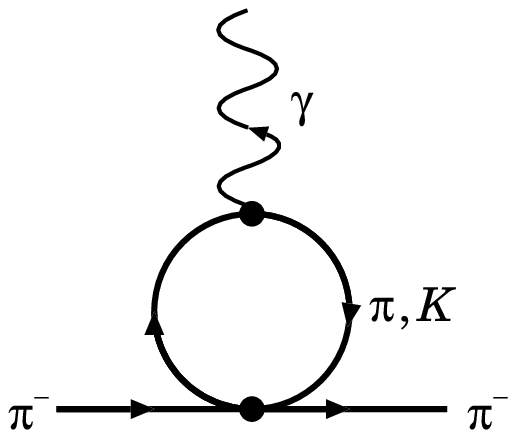}
\label{fig2-1}
}
\hspace{.5cm}
\subfigure[]{
\includegraphics*[scale = .8]{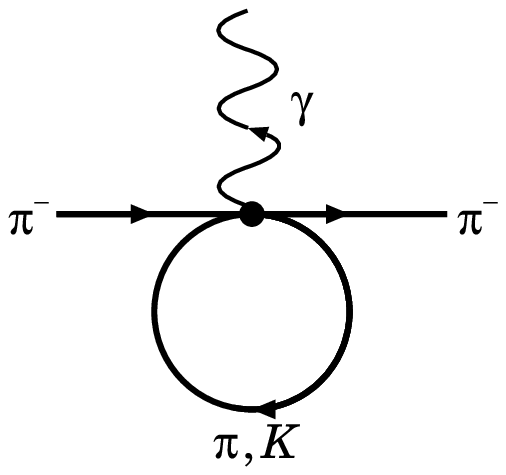}
\label{fig2-2}
}
\hspace{.5cm}
\subfigure[]{
\includegraphics*[scale = .8]{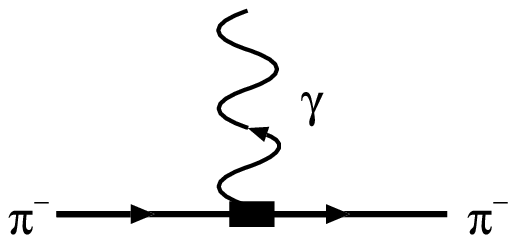}
\label{fig2-3}
}
\hspace{.5cm}
\subfigure[]{
\includegraphics*[scale = .8]{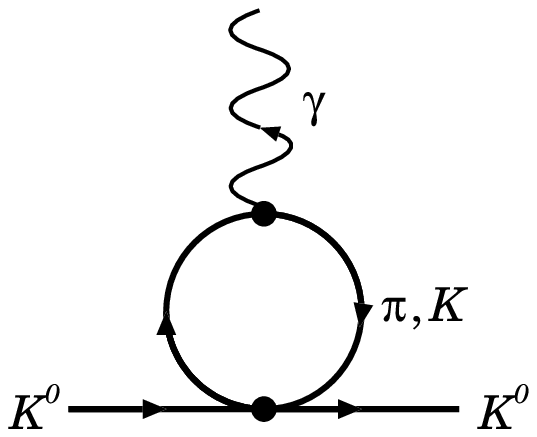}
\label{fig2-4}
}
\hspace{.5cm}
\subfigure[]{
\includegraphics*[scale = .8]{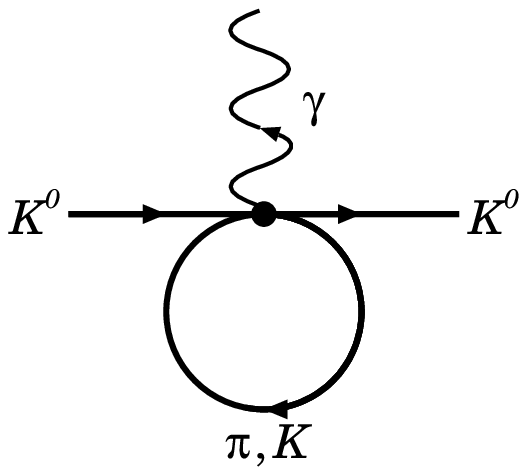}
\label{fig2-5}
}
\caption{The NLO diagrams contributing to the pion
and the neutral kaon form factors.
The dark box is NLO vertex from ${\mathcal L}^{(4)}$.}
\label{figure2}
\end{center}
\end{figure}

In $K^0_{e3e^+e^-}$ reaction, the virtual photon momentum is non-zero
and the NLO corrections to the pion and neutral kaon form factors
contribute to the reaction amplitude.
They are given by loop diagrams (Figs. \ref{fig2-1}, \ref{fig2-2},
\ref{fig2-4} and \ref{fig2-5})
and vertices from ${\mathcal L}^{(4)}$ (Fig. \ref{fig2-3}).
Only loop diagrams contribute to the neutral kaon form factor.
The amplitudes $T^{(4)}_{(a)}$ and $T^{(4)}_{(b)}$ are given as
\begin{eqnarray}
T^{(4)}_{(a)} & = &
- \frac{G_F}{\sqrt{2}}e^2V_{us}^*\frac{H^\pi(q^2)}{q^2}\left[
\bar{u}(k_2)(\slash{p_1} + \slash{p_2} + \slash{q})
(1-\gamma_5)\frac{-2p_{2 \mu }}{(p_2 + q)^2 - m_\pi ^2}
v(k_1)\right]\bar{u}(k_4)\gamma ^\mu v(k_3), \nonumber\\
& &\\
T^{(4)}_{(b)}& =& - \frac{G_F}{\sqrt{2}}e^2V_{us}^*\frac{H^K(q^2)}{q^2}\left[
\bar{u}(k_2)(\slash{p_1} + \slash{p_2} - \slash{q})
(1-\gamma_5)\frac{2p_{1 \mu }}{(p_1 - q)^2 - m_K ^2}
v(k_1)\right]\bar{u}(k_4)\gamma ^\mu v(k_3).
\nonumber \\
 & &
\end{eqnarray}
Here we define $q = p_1 - p_2$.
The form factors $H^\pi$ and $H^K$ are given as
\begin{eqnarray}
H^\pi(q^2) &=&  \frac{1}{F_0^2}\Big[
2L_9 q^2 + A(m_\pi^2) + \frac{1}{2}A(m_K^2)
- 2B_{22}(m_\pi^2 ,m_\pi^2, q^2)  - B_{22}(m_K^2,m_K^2,q^2) \Big],
\nonumber   \\
\\
H^K(q^2) &=& \frac{1}{F_0^2}[\frac{1}{2}A(m_K^2)
-\frac{1}{2}A(m_\pi^2) - B_{22}(m_K^2,m_K^2,q^2)
 + B_{22}(m_\pi^2,m_\pi^2,q^2)].
\end{eqnarray}
The functions $A(m^2)$ and $B_{22}(m_1^2,m_2^2,q^2)$ are given in
the Appendix A.

\vspace{1cm}

\subsection{$\pi K W$ vertices}

\begin{figure}[h]
\begin{center}
\subfigure[]{
\includegraphics*[scale = .8]{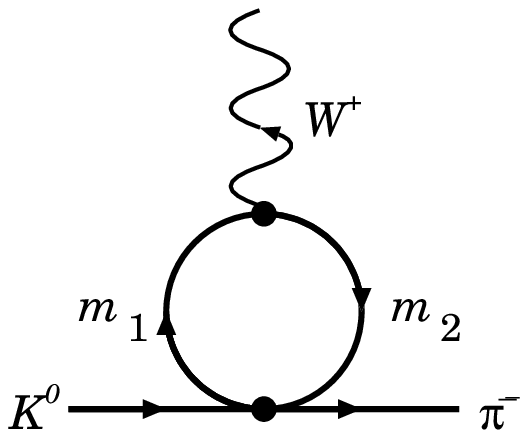}
\label{fig3-1}
}
\hspace{.5cm}
\subfigure[]{
\includegraphics*[scale = .8]{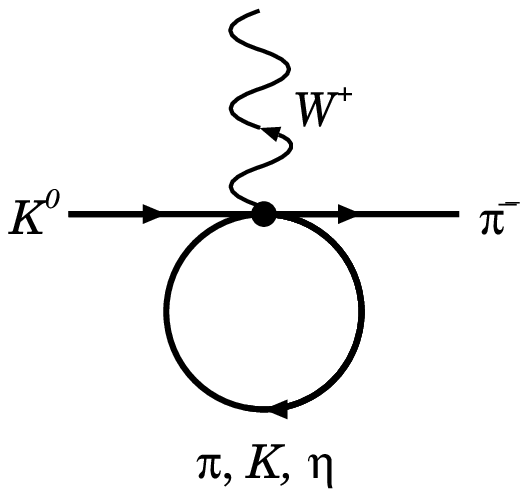}
\label{fig3-2}
}
\subfigure[]{
\includegraphics*[scale = .8]{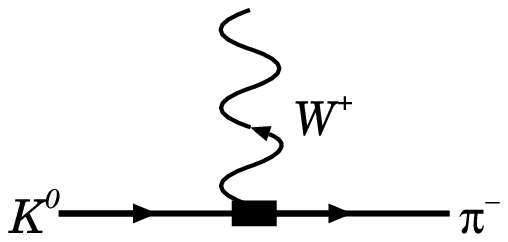}
\label{fig3-3}
}
\caption{The NLO diagrams contributing to
 $\pi K W$ vertices.
 Mesons $(m_1,m_2)$ are $(K^+,\eta)$, $(\pi^+,K^0)$, $(K^+,\pi^0)$.}
\label{figure3}
\end{center}
\end{figure}

The diagrams contributing to the NLO
 $\pi K W$ vertex are shown in Fig. \ref{figure3}.
The NLO amplitude $T^{(4)}_{(c)}$ and $T^{(4)}_{(d)}$ are given as
\begin{eqnarray}
T^{(4)}_{(c)}&=& - \frac{G_F}{\sqrt{2}}e^2V_{us}^*\frac{1}{q^2}\bigg[
\bar{u}(k_2)\left\{ G_1(r_c,l_c)(\slash{p_1} + \slash{p_2})
+ G_2(r_c,l_c)(\slash{p_1} - \slash{p_2})\right\}\nonumber \\
&&\times (1-\gamma_5)
\frac{2k_{1\mu} + \slash{q}\gamma_\mu}{(k_1 + q)^2 -m_e^2}
v(k_1)
\bigg]\bar{u}(k_4)\gamma ^\mu v(k_3),
     \\
T^{(4)}_{(d)}&=& \frac{G_F}{\sqrt{2}}e^2V_{us}^*\frac{1}{q^2}\bigg[
\bar{u}(k_2)
\left\{ G_1(r_d,l_d)\frac{\slash{p_1} + \slash{p_2} + \slash{q}}
{(p_2 + q)^2 -m_\pi ^2}
+ G_2(r_d,l_d)\frac{\slash{p_1} - \slash{p_2} - \slash{q}}
{(p_2 + q)^2 -m_\pi ^2} \right\}\nonumber\\
&& \times (1-\gamma _5) v(k_1)
\bigg]\bar{u}(k_4)2\slash{p_2} v(k_3),
\end{eqnarray}
with $r_c = p_1 + p_2, l_c = p_1 - p_2,
 r_d = p_1 + p_2 + q$ and $l_d = p_1 - p_2 - q$.
Here $G_1$ and $G_2$ are the weak form factors of the $K\pi$
transition given as
\begin{eqnarray}
G_1(r,l) &=& \frac{2L_9}{F_0^2}l^2
+ \frac{3}{8F_0^2}\left\{
A(m_\eta^2) + A(m_\pi^2) + 2A(m_K^2)
\right\}  -\frac{3}{2F_0^2}
\left\{ B_{22}(m_\pi^2,m_K^2,l^2) + B_{22}(m_K^2,m_\eta^2,l^2)\right\},
 \nonumber \\
 \\
G_2(r,l) &=&
\Big[ 4(m_K^2 - m_\pi^2)L_5 -2r\cdot l L_9 \nonumber
+\frac{1}{2}A(m_\eta^2) - \frac{5}{12}A(m_\pi^2)
+\frac{7}{12}A(m_K^2) \nonumber \\
&& \qquad  + B(m_\pi^2,m_K^2,l^2)\big\{ -\frac{1}{4}m_\pi^2
+\frac{1}{6}m_K^2 - \frac{5}{48}r^2 +\frac{5}{16}l^2
-\frac{3}{8}r\cdot l
\big\} \nonumber \\
&& \qquad  + B(m_K,m_\eta^2,l^2)\big\{ -\frac{1}{6}m_\pi^2
-\frac{1}{12}m_K^2 - \frac{1}{16}r^2 +\frac{3}{16}l^2
-\frac{3}{8}r\cdot l
\big\} \nonumber \\
&& \qquad  + B_1(m_\pi^2,m_K^2,l^2)\big\{ \frac{1}{2}m_\pi^2
-\frac{1}{3}m_K^2 + \frac{5}{24}r^2 - \frac{5}{24}l^2
+ \frac{3}{2}r\cdot l
\big\} \nonumber \\
&& \qquad  + B_1(m_K,m_\eta^2,l^2)\big\{ \frac{1}{3}m_\pi^2
+\frac{1}{6}m_K^2 + \frac{1}{8}r^2 - \frac{1}{8}l^2
+ \frac{3}{2}r\cdot l
\big\} \nonumber \\
&& \qquad  + B_{21}(m_\pi^2,m_K^2,l^2)\big\{ - \frac{5}{6}l^2
-\frac{3}{2}r\cdot l
\big\} + B(m_K,m_\eta^2,l^2)\big\{
-\frac{3}{2}r\cdot l - \frac{1}{2}l^2
\big\} \nonumber \\
&& \qquad  - \frac{5}{6}B_{22}(m_\pi^2,m_K^2,l^2)
-\frac{1}{2}B_{22}(m_K^2,m_\eta^2,l^2) \Big]\frac{1}{F_0^2}.
\end{eqnarray}
Those expressions  agree with Eqs. (4.3)-(4.4) of  \cite{bijn1} when
the kaon and the pion are on the mass shell.

\subsection{$\pi K W \gamma$ vertices}

\begin{figure}[h]
\begin{center}
\subfigure[]{
\includegraphics*[scale = .7]{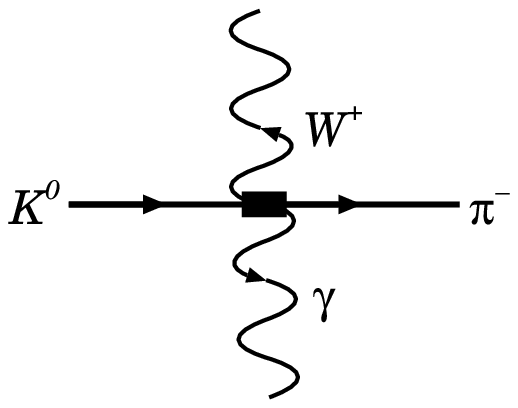}
\label{fig4-1}
}
\hspace{.5cm}
\subfigure[]{
\includegraphics*[scale = .7]{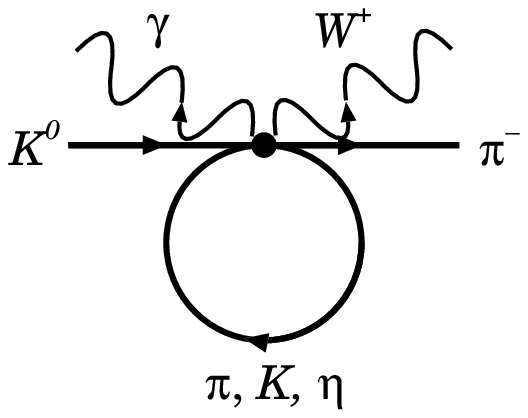}
\label{fig4-2}
}
\hspace{.5cm}
\subfigure[]{
\includegraphics*[scale = .7]{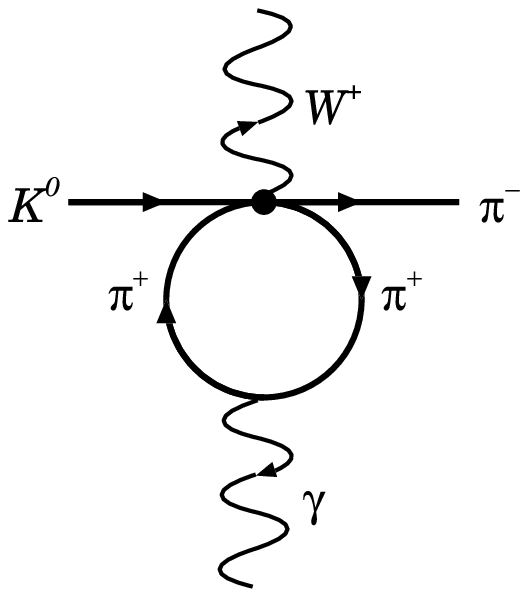}
\label{fig4-3}
}
\hspace{.5cm}
\subfigure[]{
\includegraphics*[scale = .7]{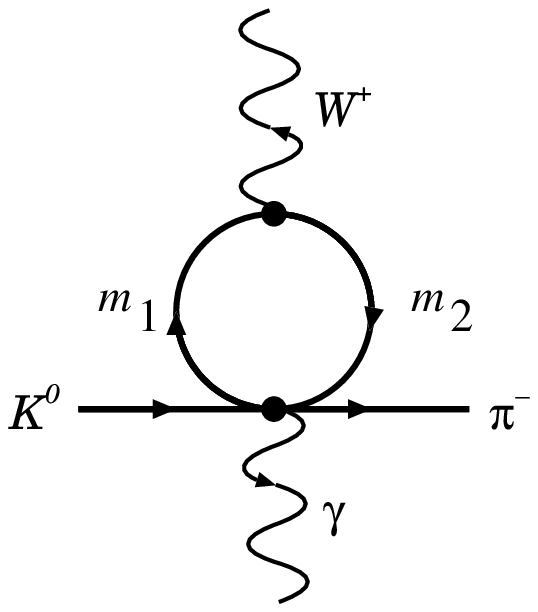}
\label{fig4-4}
}
\hspace{.5cm}
\subfigure[]{
\includegraphics*[scale = .7]{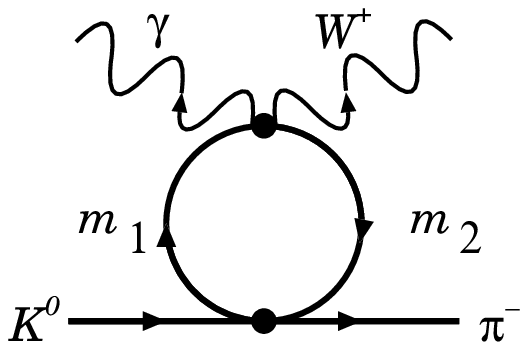}
\label{fig4-5}
}
\hspace{.5cm}
\subfigure[]{
\includegraphics*[scale = .7]{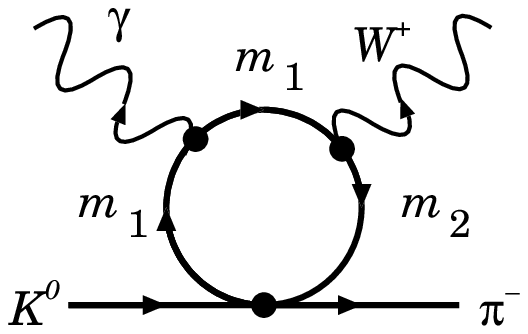}
\label{fig4-6}
}
\caption{The NLO diagrams contributing to $\pi K W \gamma$
 vertexes. Mesons $(m_1,m_2)$ are
 $(K^+,\eta)$, $(\pi^+,K^0)$, $(K^+,\pi^0)$.}
\label{figure4}
\end{center}
\end{figure}

The amplitude with the NLO correction of $\pi K W \gamma$ vertex
 is expressed as
\begin{eqnarray}
T^{(4)}_{(e)}&=& - \frac{G_F}{\sqrt{2}}e^2V_{us}^*\frac{1}{q^2}
\bar{u}(k_2)\gamma _\nu (1-\gamma _5) v(k_1)
\bar{u}(k_4)\gamma _\mu v(k_3) \sum_{\alpha = a}^{f} t^{\mu
\nu}_{(\alpha)}
\frac{1}{F_0^2}.
\end{eqnarray}
The local interaction  $\mathcal{L}^{(4)}$
shown in Fig. \ref{fig4-1} gives
\begin{eqnarray}
t^{\mu \nu}_{(a)}
&=&\
\Big[
-\frac{1}{8}A(m_\eta^2) - \frac{11}{24}A(m_\pi^2) - \frac{5}{12}A(m_K^2)
+4(m_\pi^2 - m_K^2)L_5
+4L_9(W\cdot p_1 - q\cdot p_2) \nonumber \\
& & 
+ 4L_{10}W\cdot q
\Big]g^{\mu \nu}  
+ L_9
\Big[ - 4W^\mu W^\nu - 8W^\mu p_2^\nu + 8 p_2^\mu q^\nu
+ 4p_2^\mu W^\nu - 4 q^\mu p_2^\nu  
\Big] \nonumber  \\
& &
 - 4\left[ L_9 + L_{10} \right]W^\mu q^\nu.
\end{eqnarray}
Here $q^\mu$ and $W^\mu$
are given as $q^\mu=k_3^\mu+k_4^\mu$ and $W^\mu = k_1^\mu + k_2^\mu$.

The contributions of the loop diagrams shown in
Fig. \ref{fig4-2}-\ref{fig4-6} are given as
\begin{eqnarray}
t^{\mu \nu}_{(b)}
& = & \left[
\frac{35}{12}A(m_\pi^2) + \frac{1}{4}A(m_\eta^2) + A(m_K^2)
\right]g^{\mu \nu}, \\
%%%%%%%%%%%%%%%%%%%%%%%%%%%%%%%%%%%%%%
t^{\mu \nu}_{(c)}
&=&  -\frac{10}{3} B_{22}(m_\pi^2,m_\pi^2,q^2)g^{\mu \nu},  \\
%%%%%%%%%%%%%%%%%%%%%%%%%%%%%%%%%%%%%%%
t^{\mu \nu}_{(d)}
& = &
\Bigg[
-\frac{1}{4}A(m_K^2)
- 2  B_{22}(m_K^2,m_\eta^2,W^2)
 - \frac{4}{3}B_{22}(m_\pi^2,m_K^2,W^2)
\Bigg] g^{\mu\nu}
 \nonumber \\
&+& \Bigg[
 -2 B_{21}(m_K^2,m_\eta^2,W^2)
+  \frac{1}{2} B(m_K^2,m^2_\pi , W^2) - B_1(m_K^2,m^2_\pi , W^2)
- \frac{4}{3} B_{21}(m_\pi^2,m_K^2,W^2)
\Bigg]W^\mu W^\nu
 \nonumber \\
&+&\Bigg[
       2    B_1(m_K^2 , m_\eta^2 , W^2)
 + \frac{4}{3} B_1(m_\pi^2,m_K^2,W^2)
\Bigg](p_2^\mu + 2W^\mu )W^\nu
 \nonumber \\
&+&\Bigg[
  -\frac{1}{2} B(m_K^2 , m_\eta^2 , W^2)
  -\frac{1}{3} B(m_\pi^2,m_K^2,W^2)
\Bigg](2p_2^\mu + 3W^\mu )W^\nu,
  \\
t^{\mu \nu}_{(e)}
& =  & \Bigg[
\frac{1}{2}(3p_2 + Q)\cdot QB_1(m_K^2,m_\eta^2,Q^2) 
- \left\{  p_2\cdot Q + \frac{1}{6}(m_K^2 + 2m_\pi^2) 
\right\}B(m_K^2,m_\eta^2,Q^2) 
   \nonumber  \\
& + &
   \frac{1}{4}A(m_K^2) 
 - \frac{1}{6} A(m_\pi ^2)
+ (\frac{1}{2}p_1 + p_2 + \frac{1}{3}Q)\cdot Q B_1(m_\pi^2,m_K^2,Q^2)
    \nonumber \\
&+ & 
 (\frac{1}{2}(m_\pi^2 - Q^2 ) - \frac{2}{3} p_2 \cdot Q)
   B(m_\pi^2,m_K^2,Q^2)
\Bigg] g^{\mu\nu}, \\
%%%%%%%%%%%%%%%%%%%%%%%%%%%%%%%%%%%%%%%%%%%%%%%%%%%%%%%%%%%%%%%%%
t^{\mu \nu}_{(f)}
&=& \tilde{t}^{\mu \nu} _{KK\eta} + \tilde{t}^{\mu \nu} _{KK\pi}
  +\frac{2}{3}[g^{\mu \nu} B_{22}(m_\pi^2,m_\pi^2,q^2)
      + \tilde{t}^{\mu \nu} _{\pi\pi K} ].
\end{eqnarray}
Here  $Q^\mu= q^\mu+W^\mu$.  $\tilde{t}^{\mu \nu} _{\beta}$ for
$\beta=KK\eta,KK\pi,\pi\pi K$ is defined as
\begin{eqnarray}
\tilde{t}^{\mu \nu} _{\beta} 
&=& 
- g^{\mu \nu} \left[\, a_{\beta} C_{001}(\beta)
+b_{\beta} C_{002}(\beta) + c_{\beta}C_{00}(\beta)\, \right] 
% \nonumber \\
+   p_2^\mu q^\nu  d_{\beta} 
\left[ \,C_{00}(\beta)  - C_{001}(\beta) - C_{002}(\beta)\, \right] 
 \nonumber \\
&+ &  W^\mu q^\nu \bigg[ -2C_{001}(\beta) - 4C_{002}(\beta)
 - a_{\beta}C_{112}(\beta)
- ( a_{\beta} + b_{\beta})C_{122}(\beta)
 \nonumber \\
&& - b_{\beta}C_{222}(\beta)
+ (b_{\beta} - c_{\beta})C_{22}(\beta)
+ (a_{\beta} - c_{\beta})C_{12}(\beta)
+ 2C_{00}(\beta) + c_{\beta}C_2(\beta)\, \bigg] 
\nonumber \\
&+&  W^\mu W^\nu \bigg[
-4 C_{002}(\beta) - a_{\beta}C_{122}(\beta) - b_{\beta}C_{222}(\beta)
+\left\{ \frac{1}{2}b_{\beta} - c_{\beta} \right\} C_{22}(\beta)
 \nonumber  \\
&&  + \frac{1}{2}a_{\beta}C_{12}(\beta) + C_{00}(\beta)
 + \frac{1}{2}c_{\beta}C_2(\beta) 
\, \bigg] 
\nonumber \\ 
&- &  W^\mu p_2^\nu  d_{\beta} C_{002}(\beta) 
%\nonumber \\
 + p_2^\mu W^\nu d_{\beta}
\left[\, - C_{002}(\beta) + \frac{1}{2}C_{00}(\beta)\, \right].
\end{eqnarray}
The coefficients $a_\beta,b_\beta,c_\beta,d_\beta$ and the mass parameters
$m_1^2,m_2^2,m_3^2$ of the functions
 $C_2, C_{ij},C_{ijk}$ defined in the Appendix A
are given in table I.

\begin{center}
\begin{table}
\begin{tabular}{cccc}
\hline
Coefficient  & $KK\eta$ & $KK\pi$ & $\pi\pi K$ \\
\hline
$a_\beta$  & $q\cdot (6p_2 + 2Q)$ & $2p_1 \cdot q$ &
$q\cdot (6p_2 + 2Q)$ \\ 
$b_\beta$  & $Q\cdot (6p_2 + 2Q)$ & $2p_1 \cdot Q$ &
$Q\cdot (6p_2 + 2Q)$ \\
$c_\beta$  & $-4p_2 \cdot Q-\frac{2}{3}(2m_\pi^2 + m_K^2)$ & $2(m_\pi^2 - Q^2)$ &
$-4 p_2 \cdot Q $ \\
$d_\beta$  & 6 & 2 & 6 \\
$C(m_1^2,m_2^2,m_3^2)$ 
&$C(m_K^2,m_K^2,m_\eta^2)$
&$C(m_K^2,m_K^2,m_\pi^2)$
&$C(m_\pi^2,m_\pi^2,m_K^2)$\\
\hline
\end{tabular}
\caption{Coefficients $a_\beta,b_\beta,c_\beta,d_\beta$
 and the mass parameters of the functions $C_2,C_{ij},C_{ijk}$.}
\end{table}
\end{center}

\subsection{Chiral anomaly term}

Finally the contribution of the 
chiral anomaly term $T^{(4)}_{(e,anom)}$ is given as
\begin{equation}
T^{(4)}_{(e,anom)} = - \frac{G_F}{\sqrt{2}}e^2V_{us}^*
\frac{1}{q^2}\left( - \frac{i}{8\pi^2 F_0^2} \right)
\epsilon^{\mu \nu \rho \sigma}q_\rho W_\sigma
\bar{u}(k_2)\gamma_\nu (1-\gamma_5)v(k_1)
\bar{u}(k_4)\gamma_\mu v(k_3).
\end{equation}

\section{Results and Discussions}

The total decay rate of $K^0_{e3e^+e^-}$
is given by  summing over the
spins of the leptons in the final state:
\begin{eqnarray}
\Gamma (K^0_{e3e^+e^-})
= \frac{1}{2m_K (2\pi )^{11}}
\int \, \frac{d{\bm p}_2}{2p_2^0} \,
\int \, \frac{d{\bm k}_1}{2k_1^0} \, \cdots
\int \, \frac{d{\bm k}_4}{2k_4^0}\,
\delta^4\left( p_i - p_f \right)
\sum_{f}|T_{fi}|^2. \label{10}
\end{eqnarray}
The transition matrix element $T_{fi}$ is the sum of the LO
amplitude $T^{(2)}$ and the NLO amplitude $T^{(4)}$.

The multi-dimensional phase space integration is performed by using
the Vegas integration \cite{lepage} method. One-loop integrals in
$T^{(4)}$ are evaluated numerically using the package
\textit{Looptools}\cite{hahn,oldenborgh}. In the following results,
we use  the masses of the neutral kaon and the charged pion and the charged pion
decay constant \cite{pdg}, 
\begin{equation}
m_K = 497.67 \;\textrm{MeV}, \quad
m_\pi = 139.57 \; \textrm{MeV}, \quad
F_\pi = 92.4 \; \textrm{MeV}.
\end{equation}
We use the following low energy constants at the scale of $\mu =
m_\rho = 770$ MeV from Ref.  \cite{bijnens-3},
\begin{equation}
L_9^r(m_\rho) = 6.9 \times 10^{-3}, \quad 
L_{10}^r(m_\rho) =  -5.5\times 10^{-3}
\quad  
\end{equation}
and we use $F_K/F_\pi = 1.22$ for $L_5$. The CKM matrix element and
Fermi coupling constant are chosen as $|V_{us}| = 0.220$\cite{pdg}
and $G_F = 1.16637 \times 10^{-5} GeV^{-2}$.

\begin{figure}[h]
\subfigure[]{
\includegraphics[width=7cm] {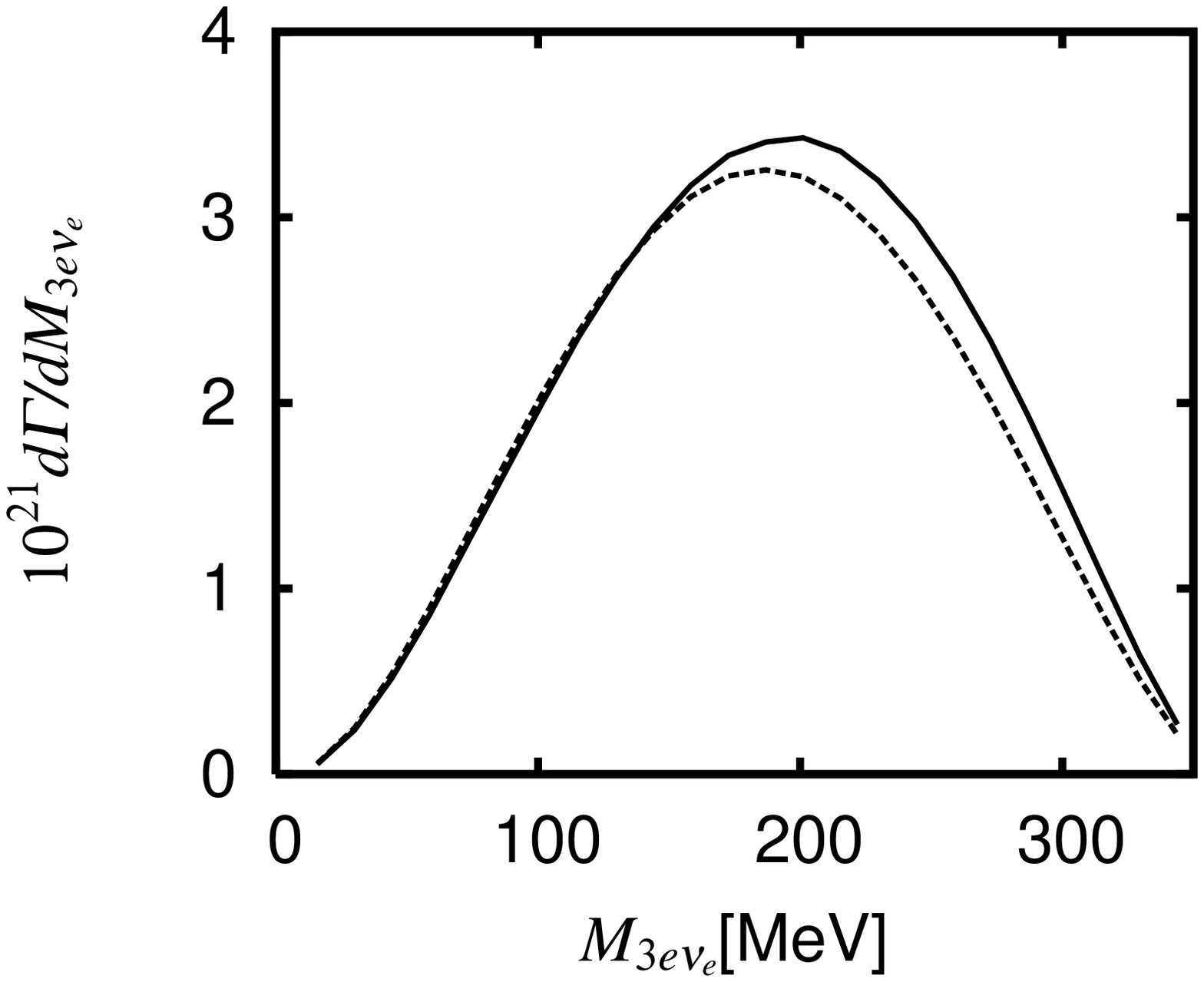}%{m3enu.ps}
\label{fig5-1}
}
\subfigure[]{
\includegraphics[width=7cm] {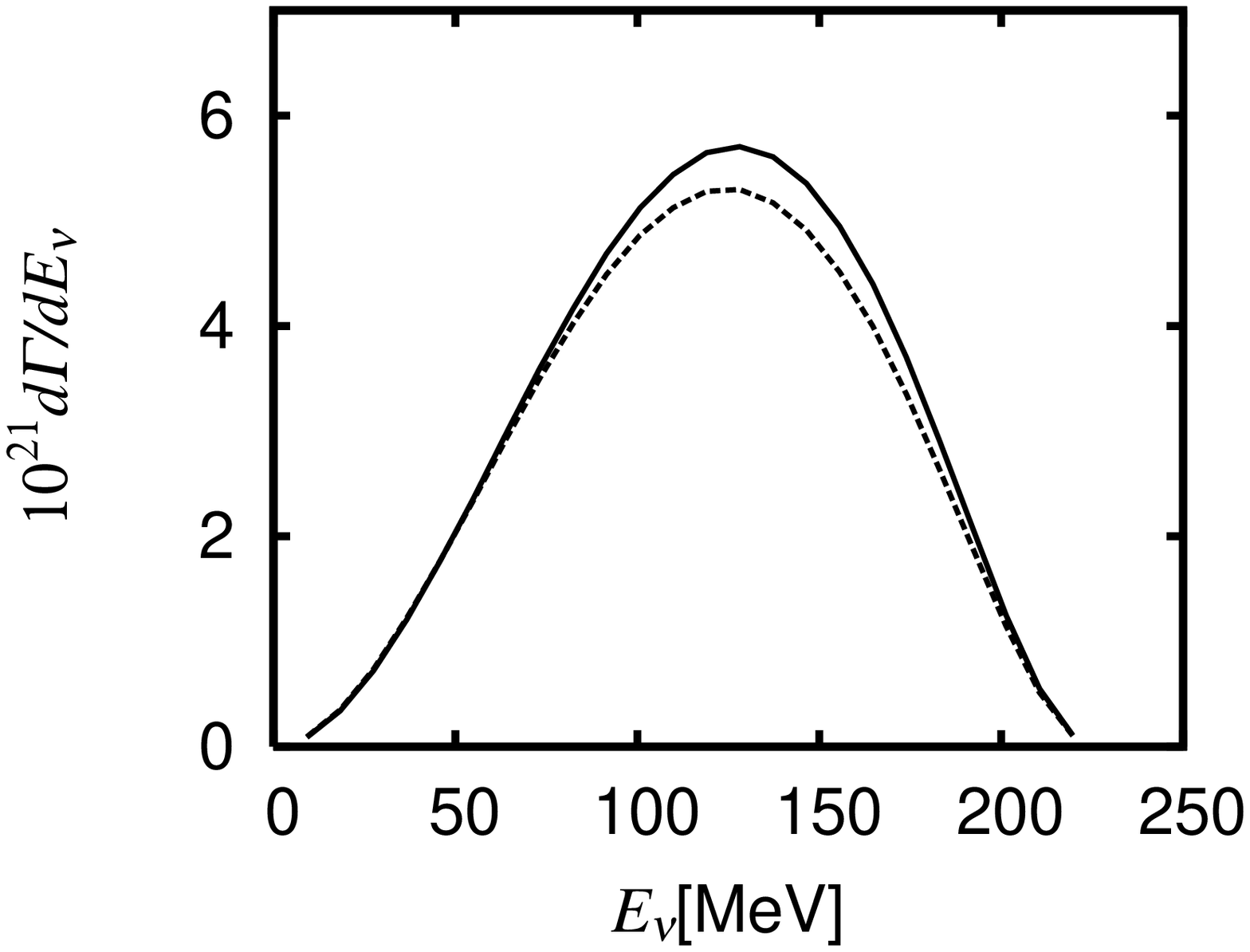}%{mnu.ps}
\label{fig5-2}
}
\subfigure[]{
\includegraphics[width=7cm] {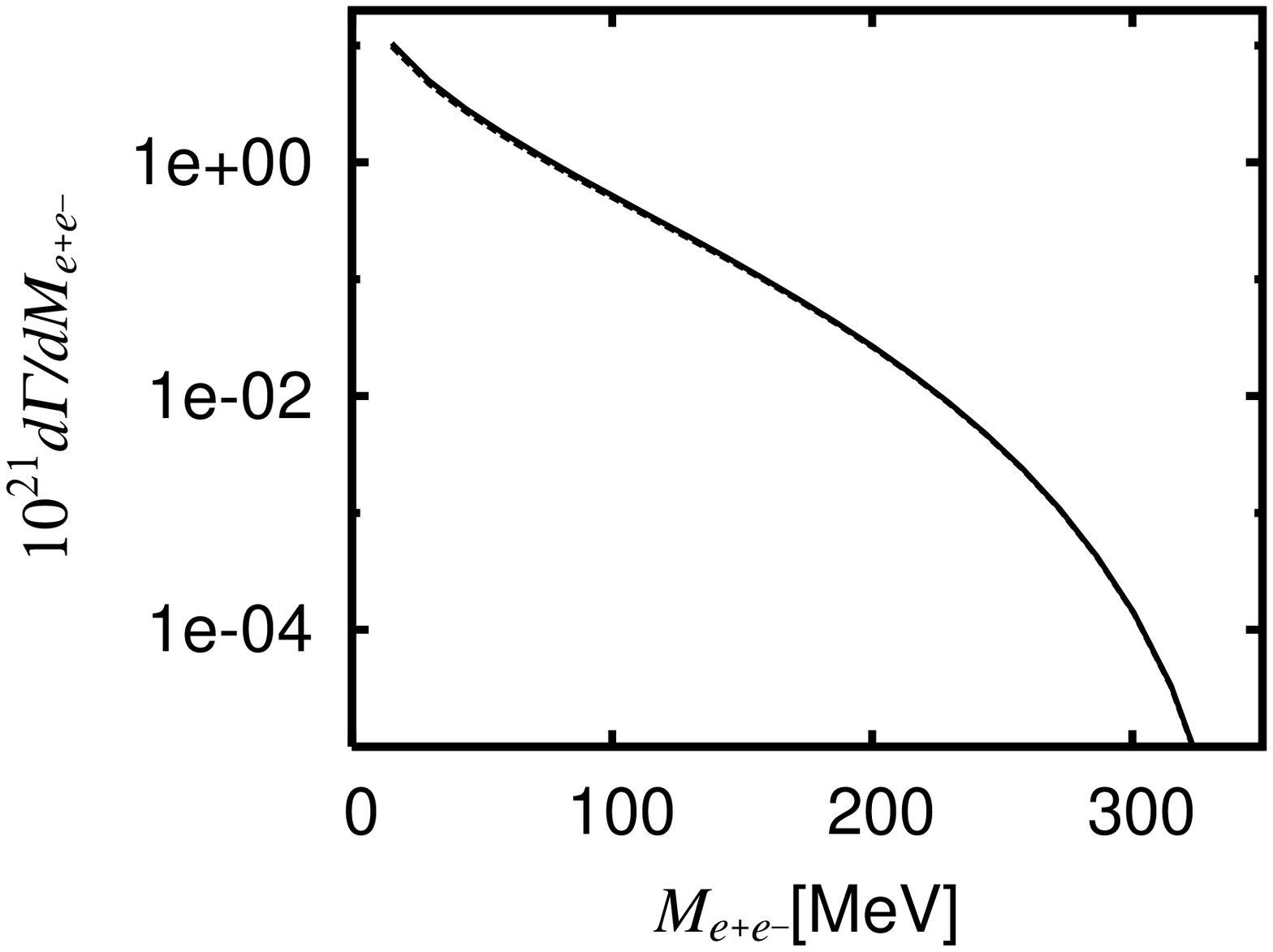}%{mee-log.ps}
\label{fig5-3}
}
\caption{
$M_{3e\nu_e}$,
$E_\nu$,
and $M_{e^+e^-}$ distributions
of $K^0_{e3e^+e^-}$ decay.
The differential decay rates are calculated with
${\mathcal O}(p^2)$( dashed curve) and
and ${\mathcal O}(p^2) + {\mathcal O}(p^4)$(solid curve).}
\label{figure5}
\end{figure}

The role of the ${\mathcal O}(p^4)$ amplitude is studied for
differential decay rates of  $K^0_{e3e^+e^-}$. We examine the energy
distribution of neutrino ($d\Gamma/dE_{\nu}$), the invariant mass
distribution of four leptons $e^+e^-e^+\nu_e$ ($d\Gamma/dM_{3e\nu}$
with $M_{3e\nu}=\sqrt{(k_1+k_2+k_3+k_4)^2}= \sqrt{(p_1-p_2)^2}$)
 and the invariant mass distribution of $e^+ e^-$
($d\Gamma/dM_{e^+ e^-}$ with $M_{e^+ e^-}=\sqrt{q^2}$). The virtual
photon momentum $q^2$ distribution is not available from
$K^0_{l3\gamma}$ decay. The calculated invariant mass distributions
$d\Gamma/dM_{3e\nu_e}$, $d\Gamma/E_\nu$ and $d\Gamma/M_{e^+e^-}$ are
shown in Fig. \ref{figure5} with the  ${\mathcal O}(p^2)$ amplitude
(dash curve) and the ${\mathcal O}(p^2)+{\mathcal O}(p^4)$ amplitude
(solid curve). In those invariant mass distributions, the second
term of Eq.(\ref{leading}) in the LO amplitude plays a dominant
role. Around the peak of those distributions, effects of the
${\mathcal O}(p^4)$ amplitude contribute about 10\% of the
$M_{3e\nu_e}$ and $E_\nu$ distributions. A smaller effect of the NLO
amplitude is found for the $M_{e^+e^-}$ distribution.

\begin{figure}[h]
\subfigure[]{
\includegraphics*[width=7cm]{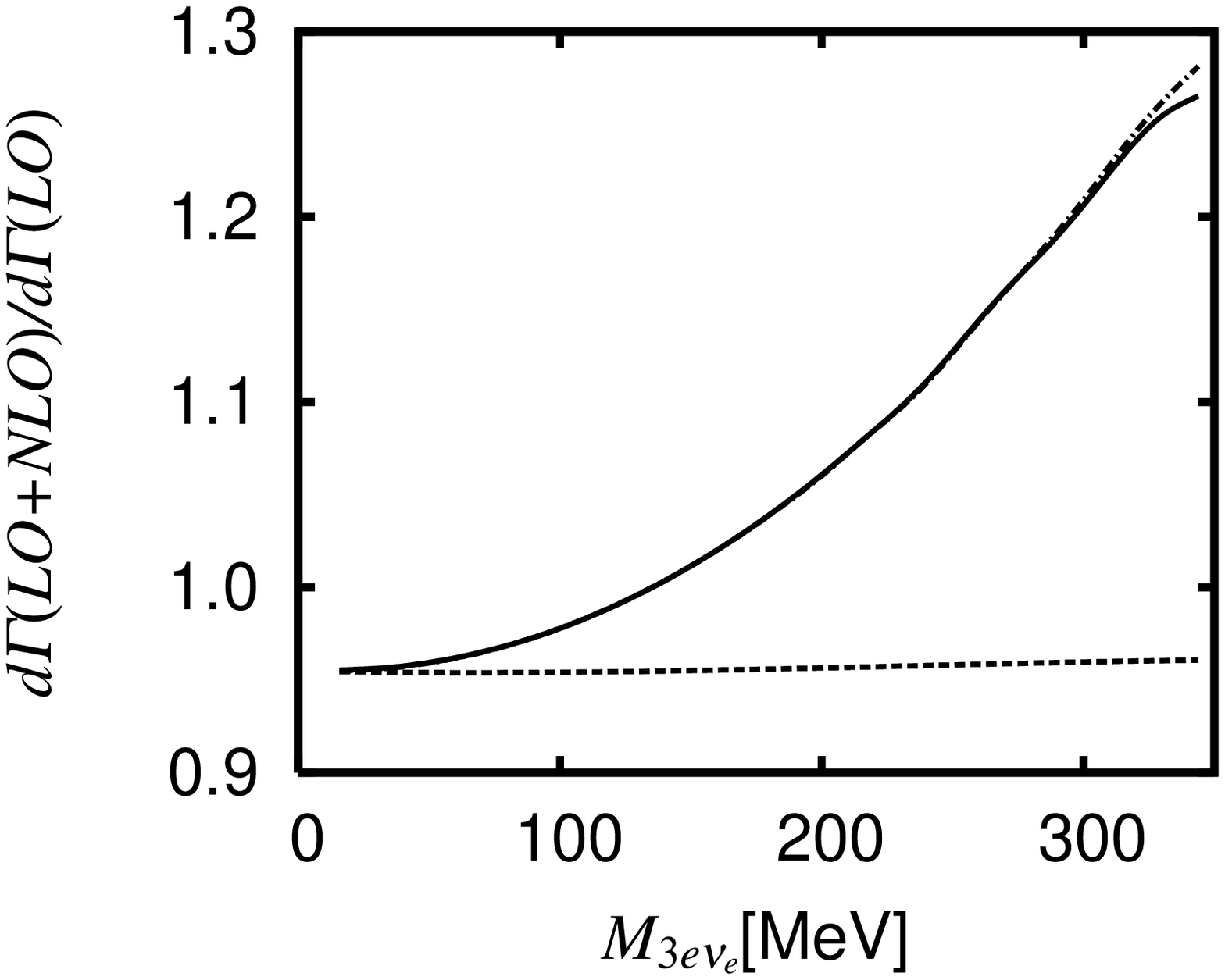}%{m3enu-ratio.ps}
\label{fig6-1}
}
\subfigure[]{
\includegraphics*[width=7cm]{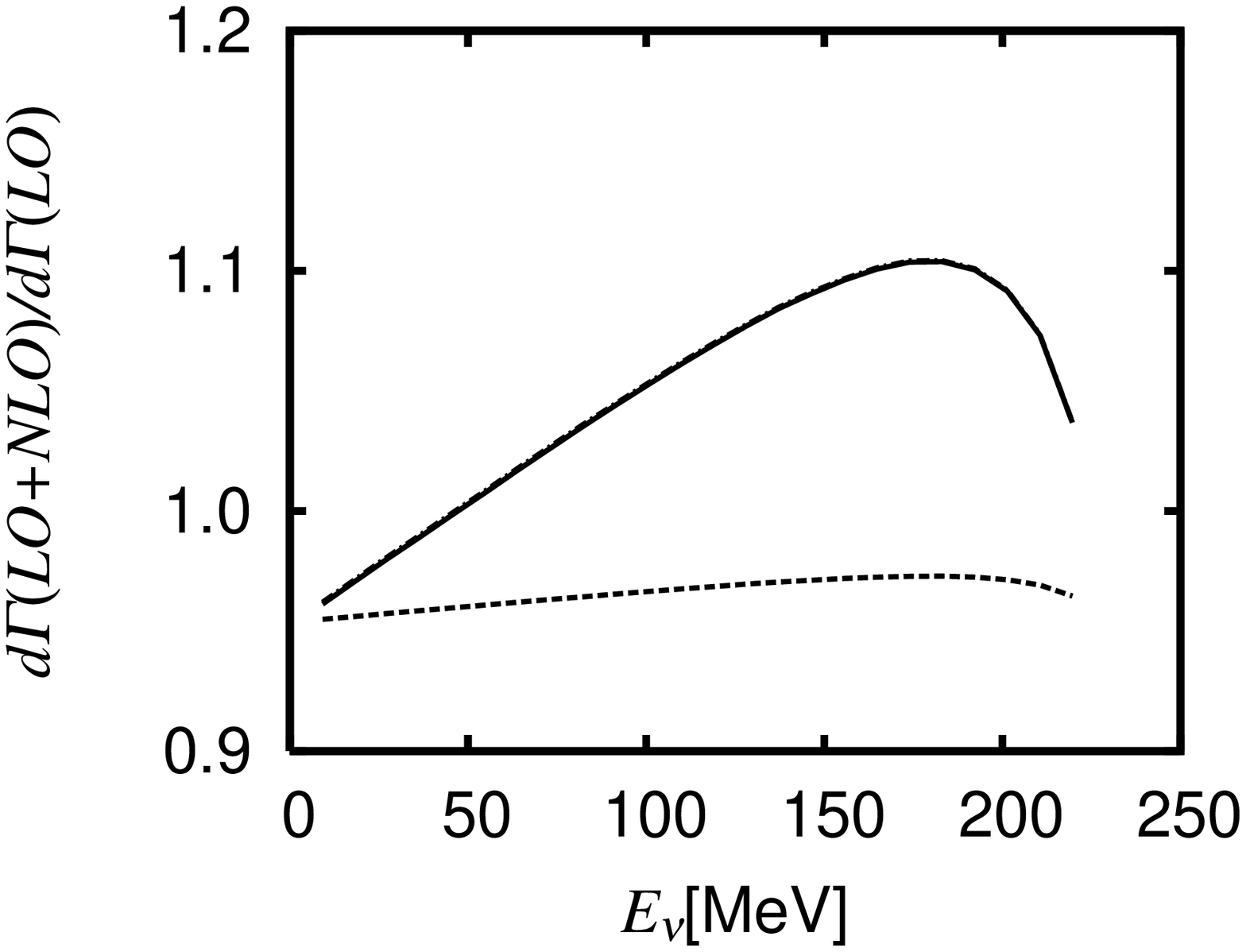}%{mnu-ratio.ps}
\label{fig6-2}
}
\subfigure[]{
\includegraphics*[width=7cm]{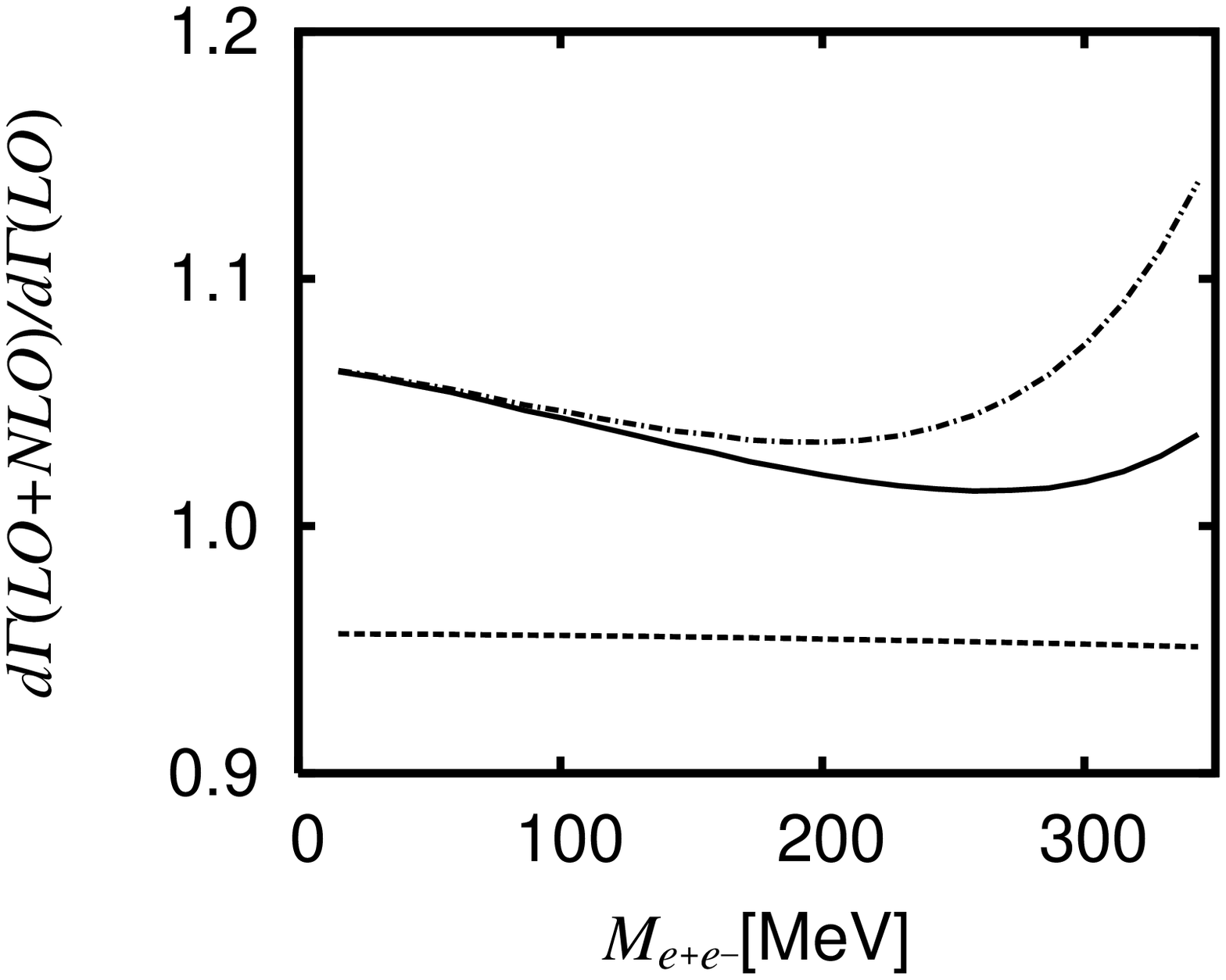}%{mee-ratio.ps}
\label{fig6-3}
}
\caption{The ratio of the $LO+NLO$ to the $LO$
for $M_{3e\nu_e}$, $E_\nu$
and $M_{e^+e^-}$  distributions.
The solid, dash-dot and dotted curves show results with
full  ${\mathcal O}(p^4)$,
${\mathcal O}(p^4)$ with $L_5^r=L_{10}^r=0$
and loop effects.
}
\label{figure6}
\end{figure}

The effect of the ${\mathcal O}(p^4)$ amplitude on the shape of the
mass distributions can be more clearly seen in the ratio
$d\Gamma(LO+NLO)/d\Gamma(LO)$. Those ratios are shown in Fig.
\ref{figure6}. The solid curves show results using the full
${\mathcal O}(p^4)$ amplitudes, while the dashed curves show results
including only loop contributions. In the dashed curves, the
${\mathcal O}(p^4)$ amplitudes are calculated with $L_5=L_{10}=0$
(dash-dot), i.e. only $L_9$ is included in addition to loop
contributions and the anomaly term. The ${\mathcal O}(p^4)$ effects
increase with energy for $M_{3e\nu_e }$ and neutrino energy
distributions. They  become about 1.2 to 1.25 for $M_{3e\nu_e} >  200$
MeV and about 10\% around the peak of the $E_\nu$ distribution. For
both  $M_{3e\nu_e }$ and $E_\nu$ distributions, it will be possible
to test the  ${\mathcal O}(p^4)$ effects in the energy region where
the decay rates themselves are large. As far as the effects of the
low energy constants are concerned, the most important contribution is $L_9$. The
effects of $L_5$,$L_{10}$ and the chiral anomaly are
 found to be small for $M_{3e\nu_e }$ and  $E_\nu$ distributions.
The loops effects reduce the invariant mass distributions by
about 5\%. The shape of the invariant mass distributions is almost
not affected by the loop diagrams. In  $M_{e^+e^-}$, the effects of
the ${\mathcal O}(p^4)$ terms appear in a slightly different way.
The $L_9$ term mainly contributes to the $M_{e^+e^-}$ distribution
below $M_{e^+e^-} = 100$ MeV. Above 150MeV, $L_{10}$ begins to
contribute and tends to cancel the contribution of $L_9$. The matrix
element of the chiral anomaly term is proportional to $\epsilon^{\mu
\nu \rho \sigma}q_\rho W_\sigma$, and the amplitude is directly
proportional to $M_{e^+e^-} = \sqrt{q^2}$. The effects of the chiral
anomaly term start to be sizable  above $M_{e^+e^-} = 150 \sim 200$
MeV and the $M_{e^+e^-}$ distribution in the high energy region will
be interesting even if the decay rate is quite small. In this energy
region, the relative importance of the 'exchange' amplitude compared
with the  'direct' amplitude increases because of the photon
propagator. A straightforward interpretation of the $q^2$ dependence
of the $M_{e^+e^-}$ distribution may be possible for
$K^0_{\mu3e^+e^-}$ decay, which is free from the exchange effects.

Finally we examine
the total decay rate of $K^0_{l3e^+e^-}(l=e,\mu)$
relative to the  $K^0_{l3}$ decay rate as
\begin{eqnarray}
{\mathcal R}(K^0_{l3e^+e^-}) 
= \frac{\Gamma (K^0_{l3e^+e^-})}
{\Gamma (K^0_{l3})}. \quad (l=e, \mu)
\end{eqnarray}
The decay rate  $\Gamma (K^0_{l3})$ is calculated in ChPT up to
${\mathcal O}(p^4)$. Here the ratios are calculated in the absence
of real and virtual photon corrections \cite{gasser-2}. Our results
on  ${\mathcal R}$ are shown in Table \ref{table1} for three cases
using the full ${\mathcal O}(p^4)$ amplitude, the ${\mathcal
O}(p^2)$ amplitude, and loop corrections. In the last case, we just
set $L_i^r = 0$. Including the ${\mathcal O}(p^4)$ amplitudes, the
total decay rate is increased by 6\%(8\%) for the
$K^0_{e3e^+e^-}$($K^0_{\mu 3e^+e^-}$) decay. The loop correction
reduces the decay rate, which is consistent with the one reported in
\cite{bijnens-2} for $K_{l3\gamma}$. The effects of the chiral
anomaly are very small on the total decay rate.

\begin{table}[h]
\begin{center}
\begin{tabular}{ccc}
\hline
\hspace*{1cm} \hspace*{1cm}
&\hspace*{1cm} ${\mathcal R}$($K^0_{e3e^+e^-}$)
&\hspace*{1cm} ${\mathcal R}$($K^0_{\mu3e^+e^-}$)\hspace*{1cm} \\
\hline
full ${\mathcal O}(p^4)$ &
$1.34 \times 10^{-4}$ &
$3.50 \times 10^{-4}$  \\
tree level &
$1.26 \times 10^{-4}$ &
$3.24 \times 10^{-4}$  \\
only loops at ${\mathcal O}(p^4)$ &
$1.20 \times 10^{-4}$ &
$ 3.12\times 10^{-4}$ \\
\hline
\end{tabular}
\end{center}
\caption{The ratio of the branching ratios of
$K^0_{e3e^+e^-}$($K^0_{\mu3e^+e^-}$) decay
to $K_{e3}$($K_{\mu3}$) decay.}
\label{table1}
\end{table}

\section{Summary}

In summary, we have studied  the differential decay rates of
$K^0_{e3e^+e^-}$ in ChPT up to ${\mathcal O}(p^4)$ for the first
time and found that the $M_{3e\nu_e }$ and  $E_\nu$ distributions
will be suitable observables to test the ${\mathcal O}(p^4)$
amplitudes. Our analysis will provide the first hint to analyze the
various mass distributions of the $K^0_{e3e^+e^-}$  decay. The data
of $K^0_{e3e^+e^-}$ from KTeV has been analyzed using the results
obtained in this work and it was found that the NLO calculation
consistently improves that of the LO one\cite{kotera}. Once the precise
data of the  $K^0_{e3e^+e^-}$  decay are available, the next task is
to separate the IB and SD contribution in this process and obtain
new information from this decay mode.

\acknowledgments
The authors would like to thank Prof. T. Yamanaka and  Dr. K. Kotera for
many useful suggestions on the analysis of KTeV data.
We also thank Prof. K. Kubodera and Drs. T. -S. H. Lee and
B. Julia-Diaz for discussions. KT was supported by the 21st Century COE
Program named "Towards a New
Basic Science: Depth and Synthesis".

\appendix

\section{Loop integrals}

Functions $A,B_{i},C_{i}$ are defined as follows.
\begin{eqnarray}
A(m_1^2)& =& \frac{\mu^{4-n}}{i}
\int \frac{d^nq}{(2\pi )^n}\frac{1}{q^2 - m_1^2},
\end{eqnarray}

\begin{eqnarray}
B(m_1^2,m_2^2,p^2) &=& \frac{\mu^{4-n}}{i}
\int \frac{d^nq}{(2\pi )^n}
\frac{1}{(q^2 - m_1^2)((q-p)^2-m_2^2)}, \\
B_\mu(m_1^2,m_2^2,p^2) &=& \frac{\mu^{4-n}}{i}
\int \frac{d^nq}{(2\pi )^n}
\frac{q_\mu}{(q^2 - m_1^2)((q-p)^2-m_2^2)} = p_\mu B_1(m_1^2,m_2^2,p^2),
\\
B_{\mu \nu }(m_1^2,m_2^2,p^2) &=& \frac{\mu^{4-n}}{i}
\int \frac{d^nq}{(2\pi )^n}
\frac{q_\mu q_\nu}{(q^2 - m_1^2)((q-p)^2-m_2^2)}\nonumber  \\
&=& p_\mu p_\nu B_{21}(m_1^2,m_2^2,p^2) + g_{\mu \nu}B_{22}(m_1^2,m_2^2,p^2),
\\
B_{\mu \nu \alpha}(m_1^2,m_2^2,p^2) &=& \frac{\mu^{4-n}}{i}
\int \frac{d^nq}{(2\pi )^n}
\frac{q_\mu q_\nu q_\alpha }{(q^2 - m_1^2)((q-p)^2-m_2^2)} \nonumber \\
&=& p_\mu p_\nu p_\alpha B_{31}(m_1^2,m_2^2,p^2)
+ (p_\mu g_{\nu \alpha } + p_\nu g_{\mu \alpha } + p_\alpha  g_{\mu \nu })
B_{32}(m_1^2,m_2^2,p^2), \nonumber \\
\end{eqnarray}

%%%%%%%%%%%%%%%%%%%%%%%%%%%%%%%%%%%%%%%%%%%%%%%%%%%%%%%%%%%%%%%%%%
\begin{eqnarray}
C(m_1^2,m_2^2,m_3^2,q^2,W^2,Q^2) &=& \frac{\mu^{4-n}}{i}
\int \frac{d^nk}{(2\pi )^n}
\frac{1}{k^2 - m_1^2}\frac{1}{(k-q)^2 - m_2^2}
\frac{1}{(k-Q)^2 - m_3^2}, \\
C_\mu (m_1^2,m_2^2,m_3^2,q^2,W^2,Q^2)
&=& \frac{\mu^{4-n}}{i} \int \frac{d^nk}{(2\pi )^n}
\frac{1}{k^2 - m_1^2}\frac{1}{(k-q)^2 - m_2^2}
\frac{1}{(k-Q)^2 - m_3^2} k^\mu \nonumber\\
&=& q_\mu C_1 + Q_\mu C_2, \\
C_{\mu \nu} (m_1^2,m_2^2,m_3^2,q^2,W^2,Q^2)  &=&
\frac{\mu^{4-n}}{i} \int \frac{d^nk}{(2\pi )^n}
\frac{1}{k^2 - m_1^2}\frac{1}{(k-q)^2 - m_2^2}
\frac{1}{(k-Q)^2 - m_3^2} k^\mu k^\nu  \nonumber\\
&=& g_{\mu \nu} C_{00}
+q_\mu q_\nu C_{11} + Q_\mu Q_\nu C_{22}
+(q_\mu Q_\nu + Q_\mu q_\nu )C_{12}, \\
C_{\mu \nu \rho} (m_1^2,m_2^2,m_3^2,q^2,W^2,Q^2)
&=&
\frac{\mu^{4-n}}{i} \int \frac{d^nk}{(2\pi )^n}
\frac{1}{k^2 - m_1^2}\frac{1}{(k-q)^2 - m_2^2}
\frac{1}{(k-Q)^2 - m_3^2} k^\mu k^\nu k^\rho \nonumber\\
&=& (g_{\mu \nu} q_\rho
+ g_{\nu \rho} q_\mu + g_{\mu \rho}q_\nu )C_{001}
+ (g_{\mu \nu} Q_\rho
+ g_{\nu \rho} Q_\mu + g_{\mu \rho}Q_\nu )C_{002} \nonumber\\
&&+(q_\mu q_\nu Q_\rho + q_\mu Q_\nu q_\rho + Q_\mu
q_\nu q_\rho )C_{112} \nonumber \\
&&+(Q_\mu Q_\nu q_\rho + Q_\mu q_\nu Q_\rho + q_\mu Q_\nu Q_\rho )
C_{122}\nonumber \\
&&+ q_\mu q_\nu q_\rho C_{111} + Q_\mu Q_\nu Q_\rho C_{222}.
\end{eqnarray}
Here $\epsilon = 4 - n$, $Q^\mu = q^\mu + W^\mu $.

\section{Comparison with the ChPT calculation of $K_{l3\gamma}$ reaction}

In the real photon limit $q^2 = 0$, one can show that our formula
for the NLO amplitudes of $K^0_{e3e^+e^-}$ agrees with the
amplitudes of $K_{l3\gamma}$ given in Ref. \cite{bijnens-2} by using
the following relations.
\begin{eqnarray}
A(m_1^2) &=& \frac{m_1^2}{16\pi^2}\lambda_0 + \bar{A}(m_1^2),\\
B(m_1^2,m_2^2,p^2) &=& \frac{\lambda_0}{16\pi^2}
+ \bar{B}(m_1^2,m_2^2,p^2), \\
B_1(m_1^2,,m_2^2,p^2) &=& \frac{\lambda_0}{32\pi^2}
+ \frac{1}{2p^2}\left\{
\bar{A}(m_2^2) - \bar{A}(m_1^2) + (m_1^2 - m_2^2 + p^2)
\bar{B}(m_1^2,m_2^2,p^2)\right\}, \\
B_{22}(m_1^2,m_2^2,p^2) &=& \frac{\lambda_0}{64\pi^2}
\left( m_1^2 + m_2^2 - \frac{p^2}{3} \right)
+ \frac{1}{96\pi^2}\left( m_1^2 + m_2^2 - \frac{p^2}{3} \right) \nonumber\\
&&+ \frac{1}{6}\bar{A}(m_2^2) + \frac{m_1^2}{3}\bar{B}(m_1^2,m_2^2,p^2)
- \frac{1}{6}(p^2 + m_1^2 - m_2^2 )\bar{B}_1(m_1^2,m_2^2,p^2), \\
B_{21}(m_1^2,m_2^2,p^2) &=& \frac{\lambda_0}{48\pi^2}
- \frac{1}{96\pi^2p^2}\left( m_1^2 + m_2^2 - \frac{p^2}{3} \right) \nonumber\\
&&+ \frac{1}{3p^2}\bar{A}(m_2^2) - \frac{m_1^2}{3p^2}\bar{B}(m_1^2,m_2^2,p^2)
+ \frac{2}{3p^2}(p^2 + m_1^2 - m_2^2 )\bar{B}_1(m_1^2,m_2^2,p^2),
\nonumber \\
\end{eqnarray}

\begin{eqnarray}
\bar{A}(m_1^2) &=& -\frac{m_1^2}{16\pi^2}\ln
\left(\frac{m_1^2}{\mu^2} \right),  \\
\bar{B}(m_1^2,m_2^2,p^2) &=& \bar{J}(p^2)
+ \frac{\bar{A}(m_1^2) - \bar{A}(m_2^2)}{m_1^2 - m_2^2}, \\
\lambda_0 &=& \frac{2}{\epsilon} + \ln (4\pi) + 1 - \gamma.
\end{eqnarray}

$\bar{J}(p^2)$ is defined in \cite{bijnens-2}.
Three point functions
$C_1,C_2,\ldots C_{222}$ can be written in rather
simple form for $q^2 = 0$.
\begin{eqnarray}
C_1 &=& \frac{Q^2 + m_1^2 - m_1^2}{2q \cdot W}C_0
+ \frac{1}{2q\cdot W}\left[ B(m_1^2,m_2^2,W^2)
-B(m_1^2,m_1^2,q^2) \right] \nonumber \\
&&-\frac{Q^2}{2(q\cdot W)^2}
\left[ B(m_1^2,m_2^2,W^2) - B(m^2_1,m^2_2,Q^2) \right],
 \\
C_2 &=& \frac{1}{2q\cdot W}\left[
B(m_1^2,m_2^2,W^2) - B(m_1^2,m_2^2,Q^2) \right],  \\
C_{00} &=& \frac{\lambda_0}{64\pi ^2} + \frac{1}{64\pi^2}
+\frac{1}{2}m_1^2C_0 + \frac{Q^2}{4q\cdot W}
\left[ \bar{B}_1(m_1^2,m_2^2,Q^2) - \bar{B}_1(m_1^2,m_2^2,W^2)
\right], \\
C_{22} &=& \frac{1}{2q\cdot W}\left[
B_1(m_1^2,m_2^2,W^2) - B_1(m_1^2,m_2^2,Q^2) \right],  \\
C_{12} &=& \frac{1}{2q\cdot W}\left[
B_1(m_1^2,m_2^2,W^2) + (Q^2 + m_1^2 - m_2^2 )C_2
-C_{00} - Q^2 C_{22} \right],  \\
C_{222} &=& \frac{1}{2q\cdot W}\left[B_{21}(m_1^2,m_2^2,W^2)
-B_{21}(m_1^2,m_2^2,Q^2) \right], \\
C_{002} &=& \frac{1}{2q\cdot W}\left[B_{22}(m_1^2,m_2^2,W^2)
-B_{22}(m_1^2,m_2^2,Q^2) \right], \\
C_{122} &=& \frac{1}{2q\cdot W}\left[B_1(m_1^2,m_2^2,W^2)
- B_{21}(m_1^2,m_2^2,W^2) \right] - \frac{1}{q\cdot W}C_{002},
\\
C_{001} &=& \frac{\lambda_0}{192\pi^2} + \frac{1}{192\pi^2}
+ \frac{1}{2}m_1^2 C_1 - \frac{1}{2}Q^2\bar{C}_{122}
+ \frac{1}{2}\bar{B}_1(m_1^2,m_2^2,W^2)
- \frac{1}{2}\bar{B}_{21}(m_1^2,m_2^2,W^2), \nonumber \\
 \\
C_{112} &=& \frac{1}{2q\cdot W}\left[
B_{21}(m_1^2,m_2^2,W^2) - 2B_1(m_1^2,m_2^2,W^2)
+B(m_1^2,m_2^2,W^2) \right]
- \frac{2}{q\cdot W}C_{001}. \nonumber \\
\end{eqnarray}


\begin{thebibliography}{99}

%\bibitem[*]{kt} Electronic address:
%   tsuji@kern.phys.sci.osaka-u.ac.jp

%\bibitem[\dag]{ts} Electronic address:
%   tsato@phys.sci.osaka-u.ac.jp


\bibitem{ktev-2}
   A. Alavi-Harati, {\it et al.}, KTeV Collaboration,
   Phys. Rev. D {\bf 64}, 112004 (2001).

\bibitem{ktev-3}
   Alexopoulos, {\it et al.}, KTeV Collaboration,
   Phys. Rev. D {\bf 71}, 012001 (2005).

\bibitem{na48-2}
   A. Lai, {\it et al.}, NA48 Collaboration,
   Phys. Lett B {\bf 605}, 247 (2005).

\bibitem{fearing-1}
E. Fishbach and J. Smith,
   Phys. Rev. {\bf 184}, 1645 (1969);\\
   H. W. Fearing, E. Fishbach and J. Smith,
   Phys. Rev. Lett. {\bf 24}, 189 (1970);\\
   H. W. Fearing, E. Fishbach and J. Smith,
   Phys. Rev. D {\bf 2}, 542 (1970).

\bibitem{holstein}
   B. R. Holstein,
   Phys. Rev. D {\bf 41}, 2829 (1990).

\bibitem{bijnens-2}
   J. Bijnens, G. Ecker and J. Gasser,
   Nucl. Phys. B {\bf 396}, 81 (1993).


\bibitem{gasser-2}
   J. Gasser, B. Kubis, N. Paver, M. Verbeni,
   Eur. Phys. J. C {\bf 40}, 205 (2005).

\bibitem{kubis}
 B. Kubis, E. H. Muller, J. Gasser and M. Schmid,
   Eur. Phys. J. C {\bf 50}, 557 (2007).

\bibitem{moulson}
 M. Moulson, hep-ex/0611057.

\bibitem{bijnens}
 J. Bijnens, hep-ph/0707.0419.

\bibitem{gasser-1}
   J. Gasser, H. Leutwyler,
   Ann. Phys. {\bf 158}, 142 (1984);\\
   J. Gasser, H. Leutwyler,
   Nucl. Phys. B {\bf 250}, 465 (1985).

\bibitem{low}
F. E. Low, Phys. Rev. {\bf 110}, 974 (1958).

\bibitem{adler}

S. L. Adler and Y. Dothan, Phys. Rev. {\bf 151}, 1267 (1966).


\bibitem{kotera}
E. Abouzaid {\it et al.}, Phys. Rev. Lett.
 {\bf 99}, 081803 (2007).


\bibitem{scherer}
   S. Scherer,
   Adv. Nucl. Phys. {\bf 27}, 277 (2002).

\bibitem{wz} 
   J. Wess and B. Zumino,
   Phys. Lett. B {\bf 37}, 95 (1971).

\bibitem{witten}
   E. Witten,
   Nucl. Phys. B {\bf 233}, 422 (1983).

\bibitem{bijn1}
   J. Bijnens and P. Talavera, Nucl. Phys. B{\bf 669}
  (2003) 341.


\bibitem{lepage}
   G. P. Lepage,
   Journal of Comp. Phys. {\bf 27}, 192 (1978).

\bibitem{hahn}
   T. Hahn and M. Perez-Victoria,
   Comput. Phys. Commun. {\bf 118}, 153 (1999).

\bibitem{oldenborgh}
   G. J. Oldenborgh and J. A. M. Vermaseren,
   Z. Phys. C {\bf 46}, 425 (1990).


\bibitem{pdg}
   S. Eidelman {\it et al.},
   Phys. Lett. B {\bf 592}, 1 (2004).

\bibitem{bijnens-3}
   J. Bijnens, G. Ecker, J. Gasser,
   hep-ph/9411232.


\end{thebibliography}
\end{document}